\documentclass[12pt,onecolumn,a4paper]{IEEEtran}

\usepackage{cite}      

\usepackage{graphicx}  

\usepackage{psfrag}    

\usepackage{subfigure} 

\usepackage{amsmath}   
\usepackage[normalem]{ulem} 
\usepackage{times,amsmath,amssymb,amsfonts}

\usepackage[usenames,dvipsnames]{pstricks}
\usepackage{epsfig}
\usepackage{pst-grad} 
\usepackage{pst-plot} 

\graphicspath{{Bilder/}}

\newcommand{\nn}{\nonumber\\ }

\newcommand{\mb}{\mathbf}
\DeclareMathOperator{\diag}{diag}
\DeclareMathOperator{\sfH}{\mathsf H}
\DeclareMathOperator{\D}{d\!}
\DeclareMathOperator{\sfT}{\mathsf T} 

\begin{document}

\title{Full Rate L2-Orthogonal Space-Time CPM for Three Antennas}

\author{Matthias~Hesse, \em{Student~Member,~IEEE,}
  J\'er\^ome Lebrun,\\ Luc~Deneire~\em{Member,~IEEE}
  \thanks{M. Hesse, L. Deneire and J. Lebrun are with the Lab. I3S, CNRS / University of
  Nice, Sophia Antipolis, France; e-mail: \{hesse,deneire,lebrun\}@i3s.unice.fr;}  \thanks{The work of M.
    Hesse is supported by the EU by a Marie-Curie Fellowship
    (EST-SIGNAL program: http://est-signal.i3s.unice.fr) under
    contract No MEST-CT-2005-021175.}\\ \vspace{1em} \texttt{\small- final version - 01.Dec.2008 -}\vspace{-2em}}


\maketitle
\thispagestyle{empty}

\begin{abstract}
  To combine the power efficiency of Continuous Phase Modulation (CPM)
  with enhanced performance in fading environments, some authors have
  suggested to use CPM in combination with Space-Time Codes (STC).
  Recently, we have proposed a CPM ST-coding scheme based on
  $L^2$-orthogonality for two transmitting antennas. In this paper we
  extend this approach to the three antennas case. We analytically
  derive a family of coding schemes which we call Parallel Code (PC).
  This code family has full rate and we prove that the
  proposed coding scheme achieves full diversity as confirmed by
  accompanying simulations. We detail an example of the
  proposed ST codes that can be interpreted as a conventional CPM scheme
  with different alphabet sets for the different transmit antennas which
 results in a  simplified implementation.   Thanks to $L^2$-orthogonality, the
  decoding complexity, usually exponentially proportional to the number 
  of transmitting antennas,  is reduced to  linear complexity.
  \end{abstract}



\section{Introduction}
\label{sec:intro}
\PARstart{T}{o} overcome the reduction of channel capacity caused by
fading, Telatar \cite{Tela99}, Foschini and Gans \cite{Fosc98}
described in the late 90s the potential gain of switching to multiple
input multiple output (MIMO) systems.  These results triggered many
advances mostly concentrated on the coding aspects for transmitting
antennas, e.g.  Alamouti \cite{Alam98} and Tarokh et al.
\cite{Taro99a} for Space-Time Block Codes (STBC) and also Tarokh et
al.\cite{Taro98} for Space-Time Trellis Codes.

Zhang and Fitz \cite{Zhan00,Zhan03} were the first to apply the idea
of STC to CPM by constructing trellis codes.  In \cite{Silv06},
Silvester et al.  derived a diagonal block ST-code which
enables non-coherent detection. In \cite{Boko07}, Bokolamulla and
Aulin described a concatenation approach to the construction
of STC for CPM. A condition for optimal coding gain
while sustaining full diversity was also recently derived by Zaji\'c
and St\"uber \cite{Zaji07}.

Inspired by orthogonal design codes, Wang and Xia introduced in
\cite{Wang04} the first orthogonal STC for two transmitting antennas
and full response CPM and later in \cite{Wang05} for partial response.
Their approach was extended in \cite{Wang03} to construct a
pseudo-orthogonal ST-coded CPM for four antennas. To avoid the
structural limitation of orthogonal design, we proposed in
\cite{Hess08a,Hess08} a STC CPM scheme based on $L^2$-orthogonality
for two antennas. Sufficient conditions for $L^2$ orthogonality were
described, $L^2$-orthogonal codes were introduced and the simulation
results displayed good performance and full rate. Here, motivated by
this results, we extend our previous work and generalize these
conditions for three transmitting antennas.

The main result of the three transmit antennas case, is that it can,
unlike the codes based on orthogonal design, achieve full diversity
with a full rate code:
 
\begin{enumerate}
\item {\bf the full rate property} is one of the main advantage of
  using the $L^2$ norm criterion, instead of merely extending the
  classical Tarokh \cite{Taro99a} orthogonal design to the CPM case.
  Indeed, in the classical orthogonal design approach, which is based
  on optimal decoding for linear modulations, the criterion is
  expressed as the orthogonality between matrices of elements, each of
  these elements being a definite integral (usually the output of a
  matched filter).  On the contrary, in the $L^2$ design approach used
  for non-linear modulations, the product in Eq.  (\ref{eq:condLong})
  is a definite integral itself, the integrand being the product of
  two signals. This allows more degrees of freedom and enables the
  full rate property.

\item {\bf the full diversity property} can be proved in a similar way
  to the classical case \cite{Zhan03}, with the help of the extensions
  proposed by Zaji\'c and St\"uber \cite{Zaji07}.
\end{enumerate}

Furthermore, it should be pointed out that the proposed coding scheme
does not limit any parameter of the CPM. It is applicable to full and
partial response CPM as well as to all modulation indexes.

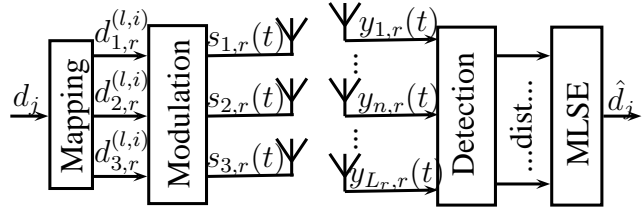
\begin{figure}
 \begin{center}
\scalebox{1} 
{
\begin{pspicture}(0,-1.35)(8.96,1.37)
\psframe[linewidth=0.04,dimen=outer](1.2529688,0.87000006)(0.65296876,-1.13)
\psframe[linewidth=0.04,dimen=outer](2.7529688,1.07)(1.9529687,-1.3299999)
\psline[linewidth=0.04cm,arrowsize=0.05291667cm 2.0,arrowlength=1.4,arrowinset=0.4]{->}(0.15296875,-0.15000005)(0.65296876,-0.15000005)
\psline[linewidth=0.04cm,arrowsize=0.05291667cm 2.0,arrowlength=1.4,arrowinset=0.4]{->}(1.2529688,0.65)(1.9529687,0.65)
\psline[linewidth=0.04cm,arrowsize=0.05291667cm 2.0,arrowlength=1.4,arrowinset=0.4]{->}(1.2529688,-0.95000005)(1.9529687,-0.95000005)
\psline[linewidth=0.04cm](2.7529688,0.65)(3.88,0.6700001)
\psline[linewidth=0.04cm](2.7529688,-0.95000005)(3.88,-0.92999995)
\usefont{T1}{ptm}{m}{n}
\rput{90.0}(0.8771874,-1.1646878){\rput(1.0209376,-0.12875015){Mapping}}
\usefont{T1}{ptm}{m}{n}
\rput{90.0}(2.2312498,-2.4756248){\rput(2.3534374,-0.13718745){Modulation}}
\psline[linewidth=0.04cm](3.8529687,-0.45000005)(3.8529687,-0.95000005)
\psline[linewidth=0.04cm](3.6529686,-0.45000005)(3.8529687,-0.75000006)
\psline[linewidth=0.04cm](3.8529687,-0.75000006)(4.052969,-0.45000005)
\psline[linewidth=0.04cm](3.8529687,1.15)(3.8529687,0.65)
\psline[linewidth=0.04cm](3.6529686,1.15)(3.8529687,0.84999996)
\psline[linewidth=0.04cm](3.8529687,0.84999996)(4.052969,1.15)
\psline[linewidth=0.04cm](4.552969,1.35)(4.552969,0.84999996)
\psline[linewidth=0.04cm](4.3529687,1.35)(4.552969,1.05)
\psline[linewidth=0.04cm](4.552969,1.05)(4.752969,1.35)
\psline[linewidth=0.04cm](4.552969,0.34999996)(4.552969,-0.15000005)
\psline[linewidth=0.04cm](4.3529687,0.34999996)(4.552969,0.04999995)
\psline[linewidth=0.04cm](4.552969,0.04999995)(4.752969,0.34999996)
\psline[linewidth=0.04cm](4.552969,-0.65000004)(4.552969,-1.15)
\psline[linewidth=0.04cm](4.3529687,-0.65000004)(4.552969,-0.95000005)
\psline[linewidth=0.04cm](4.552969,-0.95000005)(4.752969,-0.65000004)
\psline[linewidth=0.04cm,arrowsize=0.05291667cm 2.0,arrowlength=1.4,arrowinset=0.4]{->}(4.552969,0.84999996)(5.78,0.87000006)
\psline[linewidth=0.04cm,arrowsize=0.05291667cm 2.0,arrowlength=1.4,arrowinset=0.4]{->}(4.552969,-0.15000005)(5.78,-0.12999995)
\psline[linewidth=0.04cm,arrowsize=0.05291667cm 2.0,arrowlength=1.4,arrowinset=0.4]{->}(4.552969,-1.15)(5.78,-1.13)
\usefont{T1}{ptm}{m}{n}
\rput{-90.0}(4.0818753,5.129375){\rput(4.585625,0.61874986){...}}
\usefont{T1}{ptm}{m}{n}
\rput{-90.0}(5.0818753,4.129375){\rput(4.585625,-0.38125014){...}}
\psframe[linewidth=0.04,dimen=outer](6.552969,1.05)(5.752969,-1.35)
\psframe[linewidth=0.04,dimen=outer](7.9529686,1.05)(7.252969,-1.35)
\psline[linewidth=0.04cm,arrowsize=0.05291667cm 2.0,arrowlength=1.4,arrowinset=0.4]{->}(6.552969,0.65)(7.252969,0.65)
\psline[linewidth=0.04cm,arrowsize=0.05291667cm 2.0,arrowlength=1.4,arrowinset=0.4]{->}(6.552969,-1.0500001)(7.252969,-1.0500001)
\psline[linewidth=0.04cm,arrowsize=0.05291667cm 2.0,arrowlength=1.4,arrowinset=0.4]{->}(7.9529686,-0.15000005)(8.452969,-0.15000005)
\usefont{T1}{ptm}{m}{n}
\rput{90.0}(6.5875,-7.1193743){\rput(6.8334374,-0.28093725){...dist...}}
\usefont{T1}{ptm}{m}{n}
\rput{90.0}(7.344219,-7.8517184){\rput(7.597969,-0.26874956){MLSE}}
\usefont{T1}{ptm}{m}{n}
\rput{90.0}(5.965156,-6.2189064){\rput(6.0920315,-0.14187495){Detection}}
\psline[linewidth=0.04cm](2.7529688,-0.15000005)(3.88,-0.12999995)
\psline[linewidth=0.04cm](3.8529687,0.34999996)(3.8529687,-0.15000005)
\psline[linewidth=0.04cm](3.6529686,0.34999996)(3.8529687,0.04999995)
\psline[linewidth=0.04cm](3.8529687,0.04999995)(4.052969,0.34999996)
\usefont{T1}{ptm}{m}{n}
\rput(0.37,0.07500005){$d_j$}
\psline[linewidth=0.04cm,arrowsize=0.05291667cm 2.0,arrowlength=1.4,arrowinset=0.4]{->}(1.2529688,-0.15000005)(1.9529687,-0.15000005)
\usefont{T1}{ptm}{m}{n}
\rput(1.63,0.975){$d_{1,r}^{(l,i)}$}
\usefont{T1}{ptm}{m}{n}
\rput(1.63,0.17500006){$d_{2,r}^{(l,i)}$}
\usefont{T1}{ptm}{m}{n}
\rput(1.63,-0.62499994){$d_{3,r}^{(l,i)}$}
\usefont{T1}{ptm}{m}{n}
\rput(3.25,-0.72499996){$s_{3,r}(t)$}
\usefont{T1}{ptm}{m}{n}
\rput(3.25,0.07500005){$s_{2,r}(t)$}
\usefont{T1}{ptm}{m}{n}
\rput(3.25,0.87500006){$s_{1,r}(t)$}
\usefont{T1}{ptm}{m}{n}
\rput(5.27,1.075){$y_{1,r}(t)$}
\usefont{T1}{ptm}{m}{n}
\rput(5.24,-0.92499995){$y_{L_r,r}(t)$}
\usefont{T1}{ptm}{m}{n}
\rput(5.27,0.07500005){$y_{n,r}(t)$}
\usefont{T1}{ptm}{m}{n}
\rput(8.19,0.07500005){$\hat d_j$}
\end{pspicture} 
}
  \caption{Structure of a MIMO Tx/Rx system }
  \label{fig:block}
  \end{center}
\end{figure}

We first give the system model for a multiple input multiple output
(MIMO) system with $L_t$ transmitting (Tx) antennas and $L_r$
receiving (Rx) antennas (Fig. \ref{fig:block}). Later on, we will use
this general model to derive an L2-OSTC for CPM for $L_t=3$.  The
emitted signals $\mathbf s(t)$ are mixed by a channel matrix $\mathbf
A$ of dimension $L_r \times L_t$. The elements of $\mathbf A$,
$\alpha_{n,m}$, are Rayleigh distributed random variables and
characterize the fading between the $n^{th}$ Rx and the $m^{th}$ Tx
antenna. The Tx signal is disturbed by complex additive white Gaussian
noise (AWGN) with variance of $1/2$ per dimension which is represented
by a $L_r \times L_t$ matrix $ \mathbf n(t)$. The received signal
\begin{equation}
  \mathbf y(t)=\mathbf A\mathbf s(t)+\mathbf n(t).
\end{equation}
has elements $y_{n,r}$ and dimension $L_r \times L_t$. We
group the transmitted CPM signals into blocks
\begin{equation}
  \mathbf s(t)=%
  \begin{bmatrix}
    s_{1,1}(t)&\ldots&s_{1,L_t}(t)\\
    \vdots &s_{m,r}(t)&\vdots\\
    s_{L_t,1}(t)&\ldots&s_{L_t, L_t}(t)\\
  \end{bmatrix}
  \label{eq:smatrix}
\end{equation}
similar to an ST block code with the difference that now the elements
are functions of time and not constant anymore. The elements of Eq.
(\ref{eq:smatrix}) are given by
\begin{equation}
  s_{m,r}(t)=\sqrt{\frac{E_s}{L_tT}}\exp\left( j2\pi\phi_{m,r}(t)\right)
  \label{eq:defs}
\end{equation}
for $(L_tl+r-1)T\leq t\leq (L_tl+r)T$ and $m,r=1,2,\ldots,L_t$. Here
$m$ represents the transmitting antenna and $r$ the relative time slot
in the block. The symbol energy $E_s$ is normalized to the number of
Tx antennas $L_t$ and the symbol length $T$. The continuous phase
\begin{equation}
  \phi_{m,r}(t)=\theta_m(L_tl+r)+h\sum_{i=1}^{\gamma}d_{m,r}^{(l,i)}q(t-i'T)+c_{m,r}(t)
  \label{eq:defphi}
\end{equation}
is defined similarly to \cite{Ande86} with an additional correction
factor $c_{m,r}(t)$ detailed in Section \ref{sec:CorrF}.  Furthermore,
$l$ is indexing the whole code block, $i$ the overlapping symbols for
partial response and $i'=L_tl+r-i$. With this extensive description of
the symbol $d_{m,r}^{(l,i)}$, we are able to define all possible
mapping schemes (cp. Section \ref{sec:mapping}).  The modulation index
$h=2m_0/p$ is the quotient of two relative prime integers $m_0$ and
$p$ and the phase smoothing function $q(t)$ has to be continuous for
$0\leq t\leq\gamma T$, $0$ for $t\leq0$ and $1/2\leq\gamma T$. The
memory length $\gamma$ gives the number of overlapping symbols.

To maintain continuity of phase, we define the phase memory
\begin{equation}
  \theta_m(L_tl+r+1)=\theta_m(L_tl+r)+\xi_m(L_tl+r)
  \label{eq:theta}
\end{equation}
in a general way. The function $\xi(L_tl+r)$ will be fully defined in
Section \ref{sec:CorrF} from the contribution of $c_{mr}(t)$. For a
conventional CPM system, we have $c_{mr}(t)=0$ and
$\xi(2l+1)=\frac{h}{2}d_{2l+1-\gamma}$.


In Section \ref{sec:PCS}, we derive the $L^2$ conditions for a CPM
with three transmitting antennas and introduce adequate mappings and a
family of correction factors. In Section \ref{sec:propert}, we detail
some properties of the code. In Section \ref{sec:simul}, we benchmark
the code by running some simulations and finally, in Section
\ref{sec:concl}, some conclusions are drawn.

\section{Parallel Codes (PC) for 3 antennas}
\label{sec:PCS}
\subsection{$L^2$ Orthogonality}
\label{sec:l2ortho}

In this section we describe how to enforce $L^2$ orthogonality on CPM
systems with three transmitting antennas. Similarly to \cite{Hess08},
we impose $L^2$ orthogonality by
\begin{equation}
  \int\limits_{3lT}^{(3l+3)T}\mathbf s(t)\mathbf s^{\sfH}(t)\D t=E_S\mathbf I
\end{equation}
where $\mathbf I$ is the $3\times3$ identity matrix. Hence the
correlation between two different Tx antennas $s_{m,r}(t)$ and
$s_{m',r}(t)$ is canceled over a complete STC block if
\begin{equation}
  \int\limits_{3lT}^{(3l+3)T}s_{m,r}(t)s_{m',r}^*(t)\D t = 0
\end{equation}
with $m\neq m'$. Now, by using Eq. (\ref{eq:defs}) and
(\ref{eq:defphi}) we get
\begin{align}
  0=\sum_{r=1}^{3}\int\limits_{(3l+r-1)T}^{(3l+r)T}\exp\Big(&j2\pi\cdot\big[ \theta_m(3l+r)+h\sum_{i=1}^\gamma
    d_{m,r}^{(l,i)}q(t-i'T)+c_{m,r}(t)\nn
  &- \theta_{m'}(3l+r)-h\sum_{i=1}^\gamma d_{m',r}^{(l,i)}q(t-i'T)-c_{m',r}(t)\big]\Big)\D t.
  \label{eq:condLong}
\end{align}
The phase memory $\theta_m(3l+r)$ is time independent and therewith can be moved to a
constant factor in front of the integrals. Similarly to \cite{Hess08},
we introduce {\em parallel mapping} $(d_{m,r}^{(l,i)}=d_{m',r}^{(l,i)})$
for the data symbols and {\em repetitive mapping}
$(c_{m,r}(t)=c_{m,r'}(t))$ for the correction factors. The integral on
three time slots can then be merged into one time dependent factor.
Furthermore, we obtain a second, time independent factor from the
phase memory. Now, by using Eq. (\ref{eq:theta}) one can see that the
condition from Eq.  (\ref{eq:condLong}) is fulfilled if
\begin{equation}
  \label{eq:cond}
  0 = 1 + \exp(ja_1) +  \exp(ja_1) \exp(ja_2)
\end{equation}
where $a_r=2\pi\left[ \xi_m(3l+r)-\xi_{m'}(3l+r)\right]$ and we
get $-\exp(-ja_1)=1+\exp(ja_2)$. By splitting this equation into imaginary
and real parts, we have the following two conditions:
\begin{align}
  -1=&\cos(-a_1)+\cos(a_2)\\
  0=&\sin(-a_1)+\sin(a_2).
\end{align}
This system has, modulo $2\pi$, two pairs of solutions
\begin{equation}
  (a_1,a_2)\in\{(2\pi/3,2\pi/3),(4\pi/3,4\pi/3)\}.
\end{equation}

Hence $L^2$ orthogonality is achieved if $
\xi_m(3l+r)-\xi_{m'}(3l+r)=1/3$ or $ \xi_m(3l+r)-\xi_{m'}(3l+r)=2/3$
for $r=1,2$ and for all combinations of $m$ and $m'$ with $m\neq m'$.
In order to determine $\xi_m(3l+r)$, we detail in the following
section the exact mapping and the correction factor.



\label{sec:pc}

\subsection{Mapping}
\label{sec:mapping}

\begin{figure}
  \centering 
\scalebox{0.9} 
{
\begin{pspicture}(0,-3.24)(7.56,3.24)
\definecolor{color3012}{rgb}{0.4,0.4,0.4}
\definecolor{color1215}{rgb}{0.0,0.0,0.6}
\definecolor{color497}{rgb}{0.8,0.0,0.0}
\definecolor{color1000}{rgb}{0.0,0.4,0.0}
\definecolor{color1210}{rgb}{0.6,0.0,0.0}
\psframe[linewidth=0.04,linecolor=color3012,dimen=outer,fillstyle=solid](6.3,1.34)(3.9,-1.06)
\psline[linewidth=0.04cm,linecolor=color3012](4.7,1.34)(4.7,-1.06)
\psline[linewidth=0.04cm,linecolor=color3012](5.5,-1.06)(5.5,1.34)
\psline[linewidth=0.04cm,linecolor=color3012](3.9,0.54)(6.3,0.54)
\psline[linewidth=0.04cm,linecolor=color3012](3.9,-0.26)(6.3,-0.26)
\rput{-90.0}(4.96,3.84){\psframe[linewidth=0.04,dimen=outer,fillstyle=solid](5.6,0.64)(3.2,-1.76)}
\psline[linewidth=0.04cm](5.6,-0.16)(3.2,-0.16)
\psline[linewidth=0.04cm](3.2,-0.96)(5.6,-0.96)
\psline[linewidth=0.04cm](4.8,0.64)(4.8,-1.76)
\psline[linewidth=0.04cm](4.0,0.64)(4.0,-1.76)
\psframe[linewidth=0.04,dimen=outer](7.5,2.44)(0.3,1.64)
\psline[linewidth=0.04cm](1.1,2.44)(1.1,1.64)
\psline[linewidth=0.04cm](3.5,2.44)(3.5,1.64)
\psline[linewidth=0.04cm](2.7,2.44)(2.7,1.64)
\psline[linewidth=0.04cm](4.3,2.44)(4.3,1.64)
\psline[linewidth=0.04cm](6.7,2.44)(6.7,1.64)
\psline[linewidth=0.04cm](5.9,2.44)(5.9,1.64)
\psline[linewidth=0.04cm](5.1,2.44)(5.1,1.64)
\usefont{T1}{ptm}{m}{n}
\rput(5.22,-1.355){\color{color1215}$d_{3,3}^{(l,2)}$}
\usefont{T1}{ptm}{m}{n}
\rput(5.22,-0.555){\color{color1215}$d_{2,3}^{(l,2)}$}
\psframe[linewidth=0.04,dimen=outer,fillstyle=solid](4.9,-0.06)(2.5,-2.46)
\psline[linewidth=0.04cm](3.3,-0.06)(3.3,-2.46)
\psline[linewidth=0.04cm](4.1,-2.46)(4.1,-0.06)
\psline[linewidth=0.04cm](2.5,-0.86)(4.9,-0.86)
\psline[linewidth=0.04cm](2.5,-1.66)(4.9,-1.66)
\psline[linewidth=0.04cm,linestyle=dotted,dotsep=0.16cm](2.5,-0.06)(3.9,1.34)
\psline[linewidth=0.04cm,arrowsize=0.05291667cm 2.0,arrowlength=1.4,arrowinset=0.4]{->}(5.3,-2.46)(6.6,-1.16)
\psline[linewidth=0.04cm,arrowsize=0.05291667cm 2.0,arrowlength=1.4,arrowinset=0.4]{->}(2.5001307,-2.8434381)(4.8,-2.86)
\psline[linewidth=0.04cm,arrowsize=0.05291667cm 2.0,arrowlength=1.4,arrowinset=0.4]{<-}(2.1,-2.46)(2.1,-0.06)
\usefont{T1}{ptm}{m}{n}
\rput(3.62,-3.055){$r$}
\usefont{T1}{ptm}{m}{n}
\rput(1.8,-1.355){$m$}
\usefont{T1}{ptm}{m}{n}
\rput(6.25,-1.955){i}
\usefont{T1}{ptm}{m}{n}
\rput(2.92,-0.455){\color{color497}$d_{1,1}^{(l,1)}$}
\usefont{T1}{ptm}{m}{n}
\rput(2.92,-1.255){\color{color497}$d_{2,1}^{(l,1)}$}
\usefont{T1}{ptm}{m}{n}
\rput(2.92,-2.055){\color{color497}$d_{3,1}^{(l,1)}$}
\usefont{T1}{ptm}{m}{n}
\rput(3.72,-0.455){\color{color1215}$d_{1,2}^{(l,1)}$}
\usefont{T1}{ptm}{m}{n}
\rput(3.72,-1.255){\color{color1215}$d_{2,2}^{(l,1)}$}
\usefont{T1}{ptm}{m}{n}
\rput(3.72,-2.055){\color{color1215}$d_{3,2}^{(l,1)}$}
\usefont{T1}{ptm}{m}{n}
\rput(4.52,-0.455){\color{color1000}$d_{1,3}^{(l,1)}$}
\usefont{T1}{ptm}{m}{n}
\rput(4.52,-1.255){\color{color1000}$d_{2,3}^{(l,1)}$}
\usefont{T1}{ptm}{m}{n}
\rput(4.52,-2.055){\color{color1000}$d_{3,3}^{(l,1)}$}
\usefont{T1}{ptm}{m}{n}
\rput(3.62,0.245){$d_{1,1}^{(l,2)}$}
\usefont{T1}{ptm}{m}{n}
\rput(5.22,0.245){\color{color1215}$d_{1,3}^{(l,2)}$}
\usefont{T1}{ptm}{m}{n}
\rput(4.42,0.245){\color{color1210}$d_{1,2}^{(l,2)}$}
\usefont{T1}{ptm}{m}{n}
\rput(2.3,2.045){\color{color1210}\small $d_{3l+1}$}
\usefont{T1}{ptm}{m}{n}
\rput(3.9,2.045){\color{color1000}\small $d_{3l+3}$}
\psline[linewidth=0.04cm,arrowsize=0.05291667cm 2.0,arrowlength=1.4,arrowinset=0.4]{->}(0.3,2.74)(2.7,2.74)
\usefont{T1}{ptm}{m}{n}
\rput(0.91,3.045){$t$}
\usefont{T1}{ptm}{m}{n}
\rput(3.1,2.045){\color{color1215}\small $d_{3l+2}$}
\psline[linewidth=0.02cm,linestyle=dashed,dash=0.16cm 0.16cm](1.9,1.64)(2.5,-0.06)
\psline[linewidth=0.02cm,linestyle=dashed,dash=0.16cm 0.16cm](4.3,1.64)(4.9,-0.06)
\usefont{T1}{ptm}{m}{n}
\rput(1.46,2.045){\small $d_{3l}$}
\psline[linewidth=0.02cm,linestyle=dashed,dash=0.16cm 0.16cm](1.1,1.64)(3.2,0.64)
\psline[linewidth=0.02cm,linestyle=dashed,dash=0.16cm 0.16cm](3.5,1.64)(5.6,0.64)
\usefont{T1}{ptm}{m}{n}
\rput(1.32,-0.255){block of}
\usefont{T1}{ptm}{m}{n}
\rput(0.95,-0.655){data symbols}
\usefont{T1}{ptm}{m}{n}
\rput(6.51,2.745){data sequence}
\psline[linewidth=0.04cm](1.9,2.44)(1.9,1.64)
\psline[linewidth=0.04cm,linestyle=dotted,dotsep=0.16cm](5.6,0.64)(6.3,1.34)
\psline[linewidth=0.04cm,linestyle=dotted,dotsep=0.16cm](4.9,-2.36)(6.3,-1.06)
\end{pspicture} 
}
  \caption{Mapping of the data sequence to the data symbols }
  \label{fig:mapping}
\end{figure}
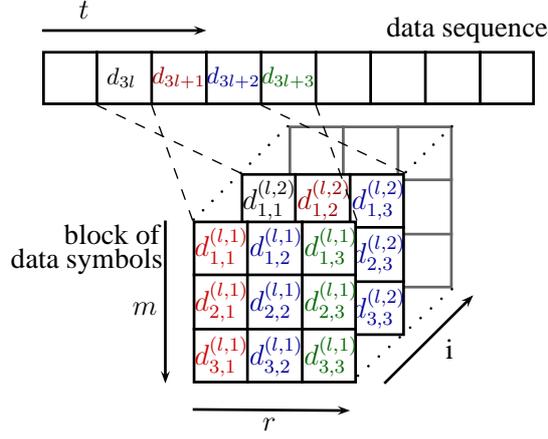

In this section we describe the mapping of the data sequence $d_j$ to
the data symbols $d_{m,r}^{(l,1)}$ of the block code (Fig.
\ref{fig:mapping}). To obtain full rate each code block has to
include three new symbols from the data sequence. In general, the
mapping of the three new symbols has no restriction. However, to construct
a mapping two criteria are considered:
\begin{itemize}
\item simplification of Eq. (\ref{eq:condLong})
\item low complexity of function $\xi_m(3l+r)$.
\end{itemize}
The first criteria is already determined by using {\em parallel
  mapping} $(d_{m,r}^{(l,i)}=d_{m',r}^{(l,i)})$. Therewith the mapping
for the $m$-dimension (Fig.  \ref{fig:mapping}) is fixed. For the
remaining two dimensions we choose a mapping similar to conventional
CPM. The subsequent data symbols in $r$-direction are mapped to
subsequent symbols from the data sequence. Also similarly to
conventional CPM we shift this mapping by $-i$ and get
\begin{equation}
  d_{m,r}^{(l,i)}=d_{3l+r-i+1}.
\end{equation}

\subsection{Correction Factor}
\label{sec:CorrF}

The choice of the phase memory and therewith of the function
$\xi_m(3l+r)$ ensures the continuity of phase. If
$\phi_{m,r}((L_tl+r)T)=\phi_{m,r+1}((L_tl+r)T)$, we always have the
desired continuity. Hence,
\begin{align}
  \xi_m(L_tl+r)=h\sum\limits_{i=1}^{\gamma}d_{m,r}^{(l,i)}q(iT)+c_{m,r}((3l+r)T)
  - h\sum\limits_{i=1}^{\gamma}d_{m,r+1}^{(l,i)}q(iT)-c_{m,r+1}((3l+r)T).
  \label{eq:xiLong}
\end{align}
With a mapping similar to conventional CPM (Section \ref{sec:mapping})
we can simplify the two sums into a single term and get for $r=1,2$
\begin{align}
  \xi_m(3l+r)=&\frac{h}{2}d_{3l+r-\gamma+1}+c_{m,r}((3l+r)T)-c_{m,r+1}((3l+r)T).
  \label{eq:xi}
\end{align}
As the data symbols are equal on each antenna, the difference between
two different $\xi_m(3l+r)$ does not depend on the data symbol
$d_{3l+r-\gamma+1}$. Thus, when choosing {\em parallel mapping}, $L^2$
orthogonality only depends on the correction factor.

To fulfill Eq. (\ref{eq:cond}) for all antennas we take
\begin{itemize}
\item for $m=1$, $m'=2$
  \begin{align}
    a_r=\frac{2\pi}{3}=&2\pi[c_{1,r}((3l+r)T)-c_{1,r+1}((3l+r)T)-c_{2,r}((3l+r)T)+c_{2,r+1}((3l+r)T)],
    \label{eq:ar11}
  \end{align}
\item for $m=2$, $m'=3$
  \begin{align}
    a_r=\frac{2\pi}{3}=&2\pi[c_{2,r}((3l+r)T)-c_{2,r+1}((3l+r)T)-c_{3,r}((3l+r)T)+c_{3,r+1}((3l+r)T)]
    \label{eq:ar12}
  \end{align}
\item and for $m=1$, $m'=3$ we consequently get
  \begin{align}
    a_r=\frac{4\pi}{3}=&2\pi[c_{1,r}((3l+r)T)-c_{1,r+1}((3l+r)T)-c_{3,r}((3l+r)T)+c_{3,r+1}((3l+r)T)].
    \label{eq:ar13}
  \end{align}
\end{itemize}
The other three possible combinations of $m$ and $m'$ with $m\neq m'$
lead only to a change of sign and we get
$a_r=-2\pi/3,-2\pi/3,-4\pi/3$, respectively. Due to the modulo $2\pi$
character of our condition, these are also valid solutions.

For simplicity, we assume similar correction factors for each time
slot $r$ of one Tx antenna $c_{m,1}(t)=c_{m,2}(t)=c_{m,3}(t)$. Since
Eq.  (\ref{eq:ar13}) arises from Eq. (\ref{eq:ar11}) and
(\ref{eq:ar12}), we have two equations and three parameters:
$c_{1,r}(t)$, $c_{2,r}(t)$ and $c_{3,r}(t)$. Hence we define
$c_{2,r}(t)=0$ for $r=1,2,3$ and we get
$c_{1,r}((3l+r)T)-c_{1,r+1}((3l+r)T)=1/3$ and
$c_{3,r}((3l+r)T)-c_{3,r+1}((3l+r)T)=-1/3$ for $r=1,2$. The codes
satisfying these conditions are coined Parallel Codes (PC). We will
now describe some possible solutions of this type.

An obvious solution for the correction factor is obtained for all
functions which are $0$ for $t=(3l+r)T$ and $\pm 1/3$ for
$t=(3l+r+1)T$, e.g.
\begin{equation}
\label{eq:c1}
  c_{1,r}(t)=-c_{3,r}(t)=\frac{2}{3}\cdot\frac{t-(3l+r)T}{2T}
\end{equation}
for $(3l+r)T\leq t\leq(3l+r+1)T$. We denote this solution as linear
parallel code (linPC). Of course, other choices are possible, e.g.
based on raised cosine (rcPC).

Another way of defining the correction factor is
\begin{equation}
\label{eq:c2}
  c_{1,r}(t)=-c_{3,r}(t)=\sum\limits_{i=1}^\gamma\frac{2}{3} q(t-i'T)
\end{equation}
for $(3l+r)T\leq t\leq(3l+r+1)T$. In that case we take advantage of
the natural structure of CPM, i.e. in Eq. (\ref{eq:xiLong}) all except
one summands cancel down, similar to the terms with the data symbols.
This definition has the advantage that we can merge the correction
factor and the data symbol in Eq. (\ref{eq:defphi}) and we obtain two
pseudo alphabets shifted by an offset (offPC) for the first and third
transmitting antenna

\begin{align}
  &\Omega_{d_1}=\left\{
    -M+1+\frac{2}{3h},-M+3+\frac{2}{3h},\ldots,M-1+\frac{2}{3h}\right\}\nn
  &\Omega_{d_3}=\left\{
    -M+1-\frac{2}{3h},-M+3-\frac{2}{3h},\ldots,M-1-\frac{2}{3h}\right\}\nonumber.
\end{align}

Consequently, this $L^2$-orthogonal design may be seen as three
conventional CPM signals with different alphabet sets $\Omega_d$,
$\Omega_{d_1}$and $\Omega_{d_3}$ for each antenna. In this method, the
constant phase offsets introduce frequency shifts. But as shown by the
simulations in next section, these shifts are quite moderate.

\section{Properties of PC CPM}
\label{sec:propert}
\subsection{Decoding}
\label{sec:decod}

In this section, for convenience, we use only one receiving antenna
but the extension to more than one is straightforward.  Hence, the
optimal receiver for the proposed codes relies on the computation of a
metric over complete ST blocks followed by a maximum-likelihood
sequence estimation (MLSE). That is, the CPM STC spans a trellis with
$pM^{\gamma-1}$ states. For non-orthogonal codes, each state leads to
$M^{L_t}=M^3$ paths with the associated distance evaluated using the $L^2$ norm
\begin{align}
 D_1= \int\limits_{3lT}^{(3l+3)T} \big|
 y_{1,r}(t)-\sum\limits_{m=1}^3 \alpha_{1,m}s_{m,r}(t) \big| ^2\D t.
 \label{eq:DISText}
\end{align}
Each metric is calculated over a length of $L_tT=3T$.

Now, from the $L^2$-orthogonality between Tx antennas, all
crosscorrelations of the symbols sent (in Eq. (\ref{eq:DISText})) are
canceled out and we get
\begin{align}
 D_2=\sum\limits_{m=1}^3
 \sum\limits_{r=1}^{3}\int\limits_{(3l+r-1)T}^{(3l+r)T}
 \big|y_{1,r}(t)-\alpha_{1,m} s_{m,r}(t)\big|^2\D t.
 \label{eq:DIST1}
\end{align}
Consequently, we are able to evaluate the metric separately for each symbol and
antenna. Therewith the number of paths reduces to $3M$ as the metric on
each path results from the sum of $3$ partial metrics by Eq.
(\ref{eq:DIST1}).

The number of paths and states can be further reduced by using some
special properties of CPM. There exist numerous efficient algorithms
for MLSE.  However, the efficiency of the detection algorithm is not
the primary scope of this paper and shall be the subject of another
upcoming paper.

\subsection{Diversity}
\label{sec:diversity}

The performance of theses codes may be evaluated using the classical
pair-wise error probability (PWEP) approach \cite{Zhan03}. Now, for
$0\leq t\leq N_cT$, let $\underline s(t)=[\underline s_1(t),\underline
s_2(t),\cdots,\underline s_{L_t}(t)]^{\sfT}$ be the vector
continuous-time representation \cite{Hess08c} of the emitted signals
\begin{equation}
  \underline
  s_m(t)= \sqrt{\frac{E_s}{L_tT}} \exp\Big(j2\pi\big[\theta_m(1) + h\sum\limits_{i=1}^{N_c}d_i q\big(t-(i-1)T\big) +c_m(t)\big]\Big).
  \label{eq:smcont}
\end{equation}
In this representation, the phase memory terms $\theta_m(L_tl+r)$ have
all been included in the summation term.  Only the initial phase
$\theta_m(1)$ remains.




We can write $\underline s(t)$ as the product of two matrices
$\underline\Theta$ and $\underline C(t)$ and a vector $\underline
d(t)$
\begin{equation}
  \underline s(t)=\sqrt{\frac{E_s}{L_tT}}~ \underline\Theta\,\underline C(t) \,\underline d(t)
  \label{eq:sVec}
\end{equation}
The matrix of initial values $\underline
\Theta=\diag(\underline\theta)$ and the matrix of correction factors
$\underline C(t)=\diag(\underline c(t))$ are $L_t\times L_t$ diagonal
matrices obtained from the vectors
\begin{align}
  \underline\theta=&\left[\begin{smallmatrix}
      \exp(j2\pi\theta_1(1))\\
      \exp(j2\pi\theta_2(1))\\
      \vdots\\
      \exp(j2\pi\theta_{L_t}(1))\\
    \end{smallmatrix}\right],& \underline
  c(t)=&\left[\begin{smallmatrix}
      \exp(j2\pi c_1(t))\\
      \exp(j2\pi c_2(t))\\
      \vdots\\
      \exp(j2\pi c_{L_t}(t))\\
    \end{smallmatrix}\right].
\end{align}
As a result of the parallel mapping, the vector of data symbols can be
written as
\begin{equation}
  \label{eq:dpm}
  \underline d(t)=\exp\Big(j2\pi h\sum\limits_{i=1}^{N_c} d_iq\big(t-(i-1)T\big)\Big)\cdot
  {\begin{bmatrix} 1 &1  & \hdots& 1
    \end{bmatrix}}^{\sfT}.
\end{equation}

To prove that our ST-Codes achieve full diversity, it is sufficient to
show that their signal matrix $\mb C_s$ has full rank
\cite{Zhan03,Zaji05}

\begin{equation}
  \mb C_s=\left[\begin{smallmatrix}
      \int\limits_0^{N_cT}|\Delta_1(t)|^2\D t& \int\limits_0^{N_cT}\hspace{-0.5em}\Delta_1(t)\Delta_2^*(t)\D t&\cdots& \int\limits_0^{N_cT}\hspace{-0.5em}\Delta_1(t)\Delta_{L_t}^*(t)\D t \\
      \int\limits_0^{N_cT}\hspace{-0.5em}\Delta_2(t)\Delta_1^*(t)\D t& \int\limits_0^{N_cT}|\Delta_2(t)|^2\D t &\cdots& \int\limits_0^{N_cT}\hspace{-0.5em}\Delta_2(t)\Delta_{L_t}^*(t)\D t \\
      \vdots&&&\vdots\\
      \int\limits_0^{N_cT}\hspace{-0.5em}\Delta_{L_t}(t)\Delta_1^*(t)\D t& \int\limits_0^{N_cT}\hspace{-0.5em}\Delta_{L_t}(t)\Delta_2^*(t)\D t&\cdots& \int\limits_0^{N_cT}|\Delta_{L_t}(t)|^2\D t \\
    \end{smallmatrix}\right]
  \label{eq:cs}
\end{equation} 
where the normalized vector $\mb\Delta(t)$ gives the difference
between the truly sent signals $\underline s_m(t)$ and the estimated
ones $\tilde{\underline s}_m(t)$:
\begin{equation}
  \mb\Delta(t)=\left[\begin{smallmatrix}
      \Delta_1(t)\\ \Delta_2(t)\\ \vdots \\ \Delta_{L_t}(t)
    \end{smallmatrix}\right]= \sqrt{\frac{L_tT}{E_s}}
  \left[\begin{smallmatrix}
      \underline s_1(t)-\tilde{\underline s}_1(t)\\ \underline s_2(t)-\tilde{\underline s}_2(t)\\ \vdots \\
      \underline s_{L_t}(t)-\tilde{\underline s}_{L_t}(t)
    \end{smallmatrix}\right].
  \label{eq:Delta}
\end{equation}
Now, Zhang and Fitz proved in \cite[Prop. 1]{Zhan03} that a necessary
and sufficient condition for $\mb C_s$ to be full rank is to have $\mb
u^{\sfH} \mb\Delta(t) \neq 0$ for all $\mb u \in \mathbb C^{L_t}$
unless $\mb u = \mb 0$.  From Eq.  (\ref{eq:sVec}), we have
\begin{equation}
  \label{eq:uDelta}
  \mb u^{\sfH} \mb\Delta(t) = \mb u^{\sfH} \underline\Theta\,
  \underline C(t)\, \big(\underline d(t) -\tilde{\underline d}(t)\big).
\end{equation}
By introducing the scalar function
\begin{align}
  \varepsilon_{(\mb d,\tilde{\mb d})}(t) =& \exp\Big(j2\pi h\sum\limits_{i=1}^{N_c}
    d_iq\big(t-(i-1)T\big)\Big)-\nn
  &\exp\Big(j2\pi h\sum\limits_{i=1}^{N_c}\tilde{d}_iq\big(t-(i-1)T\big)\Big)
\end{align}
we get from Eq.(\ref{eq:dpm}) (parallel mapping) that
\begin{equation} \mb u^{\sfH} \mb\Delta(t) = \varepsilon_{(\mb
    d,\tilde{\mb d})}(t) \mb u^{\sfH} \underline\Theta\, \underline
  C(t)\, {\begin{bmatrix} 1 &1 & \hdots& 1
    \end{bmatrix}}^{\sfT}
\end{equation}
Then, by rewriting Eq. (\ref{eq:c1}) and (\ref{eq:c2}) as
$c_m(t)=\frac{m-1}{L_t}\bar c(t)$,
\begin{equation}
 \mb u^{\sfH} \mb\Delta(t) = \varepsilon_{(\mb d,\tilde{\mb d})}(t)
\sum\limits_{m=1}^{L_t} u_m^* \exp\Big(j2\pi\big[\theta_m(1)+\frac{m-1}{L_t}\bar c(t)\big]\Big)
\end{equation}
Since for $\mb d\neq \tilde{\mb d}$, $ \varepsilon_{(\mb d,\tilde{\mb d})}(t)\neq 0$, to get $ \mb u^{\sfH} \mb\Delta(t) = 0$ it would imply that
\begin{equation}
  \label{eq:udel_null}
  \sum\limits_{m=1}^{L_t} \big(u_m^*\exp(j2\pi\theta_m(1))\big) \exp(j2\pi \frac{m-1}{L_t}\bar c(t)) = 0. 
\end{equation}
Introducing the polynomial
\begin{equation}
  p(x) = \sum\limits_{m=1}^{L_t}
  \big(u_m^*\exp(j2\pi\theta_m(1))\big)\, x^{m-1},
\end{equation}
Eq. (\ref{eq:udel_null}) means that $p(e^{j2\pi \bar c(t)/L_t}) = 0 $.
From the definition of $\bar c(t)$ (i.e. Eq. (\ref{eq:c1}) and
(\ref{eq:c2})), this would imply that the polynomial $p(x)$ of degree
$L_t-1$ vanishes on more than $L_t$ different points. Thus, $p\equiv 0$
and $u_m = 0$ for all $m$.  Consequently, by \cite[Prop.~1]{Zhan03},
the signal matrix $\mb C_s$ has full rank and all the codes (linPC and
OffPC) achieve full diversity.

\section{Simulations}
\label{sec:simul}
In this section we test the proposed algorithms by running MATLAB
simulations. More precisely, we benchmark the offPC and linPC codes for two and
three Tx antennas. For all simulations we use a Gray-coded CPM with a
modulation index of $1/2$, an alphabet of 2 bits per symbol $(M=4)$
and a memory length $\gamma$ of $2$. We use a linear phase smoothing
function $q(t)$ (2REC) and for the initial phase $\theta_i(1)$ optimal
values corresponding to \cite{Hess08c} are used.

The modulated signals are transmitted over a frequency flat Rayleigh
fading channel with complex additive white Gaussian noise. The fading
coefficients $\alpha_{n,m}$ are constant for the duration of a code
block (block fading) and known at the receiver (coherent detection).
To guarantee a fair treatment of single and multi antenna systems the
fading has to have a mean value of one.

\begin{figure}
  \centering
  \includegraphics[width=0.6\columnwidth]{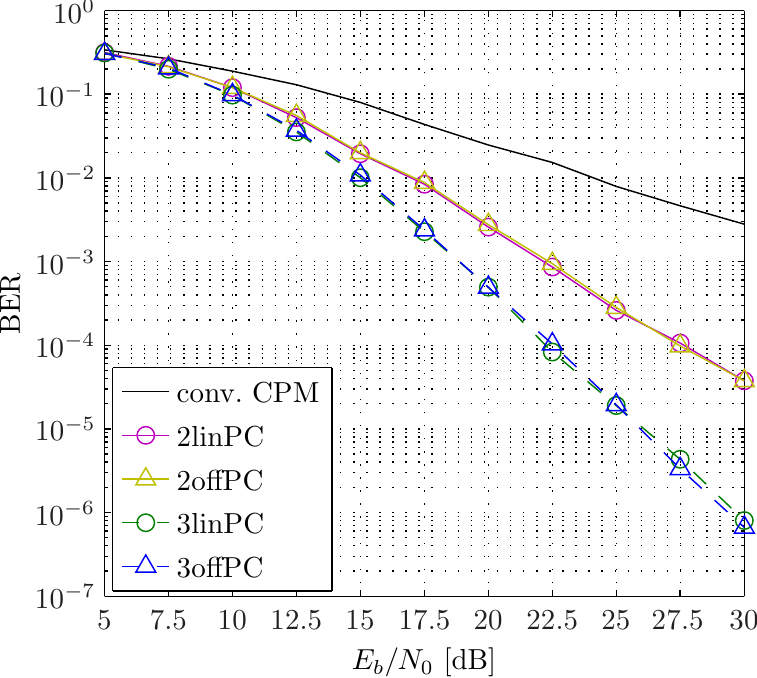}
  \caption{Simulated bit error rate (BER) over a Rayleigh fading
    channel }
  \label{fig:BER}
  \vspace{-1.5em}
\end{figure}

The received signal $y_n(t)$ is demodulated by the method introduced
in Section \ref{sec:decod}. The evaluated distances are fed to the
Viterbi algorithm (VA), which we use for MLSE. The trellis decoded by
the Viterbi algorithm has $pM^{\gamma-1}=16$ states and $L_tM=12$
paths leaving each state. In our simulations, the Viterbi algorithm is
truncated to a path memory of 10 code blocks, which means that we get
a decoding delay of $3\cdot 10T$.

Figure \ref{fig:BER} shows the simulations results for one, two and
three transmitting antennas. Both, linPC and offPC perform equally and
corresponding to Section \ref{sec:diversity} it can be seen that full
diversity is achieved. For high SNR the BER decreases with 5dB/dec and
the three Tx antennas code achieves a decay of some 3.5dB/dec.


One of the main reason for using CPM for STC is their spectral
efficiency. It should be noticed nevertheless that the introduction of
a correction factor with non-zero mean in the phase signal induces a
bandwidth expansion and thus degrades this spectral efficiency.
However, as detailed in \cite{Hess08c} the additional bandwidth
requirements are quite moderate. Namely, to achieve an attenuation in
power of -30dB for a linPC code with 3 Tx antennas, $h=1/2$, $g=2$ and
$M=8$, a relative bandwidth expansion of only 7\% is sufficient.


\section{Conclusion}
\label{sec:concl}

In this paper, we introduce a new family of $L^2$-orthogonal STC for
three antennas. These systems are based on CPM supplemented by
correction factors to ensure $L^2$-orthogonality. Structurally the
proposed code family has full rate and we provide the proof of full
diversity.  Furthermore, we detail two simple representatives of the
code family (offPC, linPC), where the offPC offers better performance
and a very intuitive representation. Finally, it should be noticed
that our approach can easily be generalized to any number of
transmitting antennas with full rate codes that also achieve full
diversity.







\bibliographystyle{IEEEtran}


\bibliography{IEEEabrv,bib}
\end{document}